%% file: asteroid_eclipse.tex
\documentclass[preprint2,10pt,letterpaper]{aastex}

\usepackage{times}

\usepackage{psfrag}

\usepackage[dvips]{color}
\definecolor{notetextcolor}{rgb}{0,.7,0}

\newcommand{\apo}{(99942)~Apophis}

\title{Eclipse Prediction and Orbit Improvement for Asteroids: \\ Theory and Application to Near Earth Asteroids}

\author{P.~Tricarico$^{1,2,*}$, N.~C.~Hearn$^{1}$, G.~Lake$^{1,3}$ and G.~Worthey$^{1}$}
\affil{$^{1}$Department of Physics and Astronomy, Washington State University, Pullman, Washington}
\affil{$^{2}$Planetary Science Institute, Tucson, Arizona}
\affil{$^{3}$Institute for Theoretical Physics, University of Zurich, Zurich, Switzerland}
\affil{$^{*}$Corresponding author. E-mail: \url{tricaric@psi.edu}}

\begin{document}

\begin{abstract}
Asteroids can be eclipsed by other bodies in the Solar System,
but no direct observation of an asteroid eclipse has been reported to date.
We describe a statistical method to predict an eclipse for an
asteroid based on the analysis of the orbital elements covariance matrix.
By propagating a set of Virtual Asteroids to an epoch
correspondent to a close approach with a Solar System planet or natural satellite, 
it is possible to estimate the probability of a partial or total eclipse.

The direct observation of an eclipse can provide data useful to improve
the asteroid orbit, especially for dim asteroids typically
observed only for a few days.
We propose two different methods: the first,
based on the inclusion of the apparent magnitude residuals into the 
orbit's least squares minimization process,
capable of improving the 
asteroid's nominal orbit and the related covariance matrix; 
the second, based on weighting different Virtual Asteroids
in relation to their apparent magnitude during the eclipse,
useful for recovery purposes.

As an application, we have numerically investigated the possibility of a  
Near Earth Asteroid eclipsed by the Moon or the Earth
in the 1990-2050 period.
A total of 74 distinct eclipses have been found, involving 59 asteroids.
In particular, the asteroid \apo{} has a probability of about 74\%
to enter the Moon's penumbra cone and a probability of about 6\%
to enter the umbra cone on April 14, 2029, 
less than six hours after a very close approach to Earth.
\end{abstract}


\section{Introduction}

Near Earth Asteroids, or NEAs, have received increased attention by the scientific community and the public at large in recent years.  Over the past decade, the rate of discovery for NEAs has increased by an order of magnitude to the current rate of roughly 500 NEAs per year \citep{jpl_neo_stats}. 

In this paper, we provide a catalog of past and future asteroid eclipses, 
and propose to use eclipses of asteroids to further constrain their orbits.  
Here, an eclipse refers to a period when an NEA passes through the shadow of the Earth or the Moon.  Simply determining whether an eclipse occurs allows one to accept or reject many possible orbits for an asteroid.  Observing the decrease in luminosity due to its travel through the umbra or penumbra cones would add to the precision of its position in the sky.  A total eclipse (passage through the umbra cone) would place significant limits on the distance to the asteroid.

Observing the eclipse of an asteroid is not a trivial matter, even if one has good knowledge of the asteroid's position.  Eclipses of an asteroid by the Moon's shadow can occur at any lunar phase.  Depending on the solar elongation of the asteroid during the eclipse, the change in luminosity may be visible from an entire hemisphere on Earth, or may not be observable anywhere on Earth.  

The prediction of such events requires a statistical treatment of the accuracy of the orbital elements.  The approach used here involves sampling the most likely set of orbital elements, based on existing observations, and then simulating the dynamics of objects on these orbits with passing time.  We also need a good understanding of the factors contributing to the changes in apparent magnitude during the eclipse.

The plan of the paper is as follows.  
In Section~2, we introduce the tools needed to describe the eclipse of an asteroid.  
In Section~3, we apply these tools to the case of NEAs eclipsed by the Earth and the Moon. 
In Section~4, we estimate the eclipse probability for NEAs.
Conclusions follow in Sections~5.

\section{Statistical Eclipse Prediction}

Our statistical method involves the use of Virtual Asteroids (VAs, see \cite{milani_virtual_asteroids}) in conjunction with the concept of an Eclipse Plane (EP) to determine the probability of an eclipse. 
The theory is very similar to NEAs impact prediction, with
eclipse cones playing the role of the target body,
and the EP replacing the Target Plane (TP, see \cite{milani_virtual_asteroids}).

\subsection{Eclipse Geometry\label{sec_eclipse_geom}}

Given a spherical body 
with a radius $r$, at a distance $D$ from the Sun,
the semi-opening angle for the \emph{penumbra} and the \emph{umbra} cones are,
respectively
$\alpha_{p,u} = \arcsin((r \pm R)/D)$,
with $R$ equal to the Sun's radius, 
the ``$+$'' sign for $\alpha_{p}$ and the ``$-$'' sign for $\alpha_{u}$.
The radius $q_{p,u}$ of each section of the cone at a distance $d$ from the
center of the spherical body and along the Sun-body line is
$q_{p,u} = (r + d \sin(\alpha_{p,u}))/\cos(\alpha_{p,u})$.
When $r<R$, as is always the case in the Solar System,
$\alpha_{p} > 0$ and $\alpha_{u} < 0$,
and the umbra cone has a finite length $l = r / \sin(-\alpha_{u})$.
The Earth's umbra cone length ranges from 3.55~LD to 3.67~LD (LD~=~Lunar Distance~=~384400~km),
while for the Moon it is in the range between 0.96~LD and 1.00~LD,
depending on the distance between these objects and the Sun.
In our computations we assume $R=6.95\times 10^5$~km.
The cones are aligned on the Sun-body line,
keeping into account the light-time delay.

\subsection{Virtual Asteroids \label{va_section}}

In order to predict an eclipse for an asteroid,
we need a statistical description of the asteroid's dynamics.
This statistical description is necessary because the asteroid's
orbital elements are determined 
only up to some uncertainty.
The information on the uncertainty on each orbital element
and on the correlation between orbital elements 
is provided by the covariance matrix,
as a result of a least-squares minimization iterative process.
The covariance matrix allows the generation of an arbitrary number of Virtual Asteroids,
all within the orbital elements uncertainty.
The major advantage of sampling the asteroid orbital uncertainty
with VAs is that each VA has the same probability
to represent the asteroid, and this greatly simplifies
a statistical interpretation of the asteroid dynamics
related to the eclipse. 
The VAs provide the statistical description we need.
An introduction to VAs and 
their use can be found in \cite{milani_virtual_asteroids}.

For the analysis described in Section~\ref{sec_application} we generate VAs
using \emph{Principal Components Analysis} technique \citep{bib_PCA},
also referred to as the Monte Carlo method in the literature.
In particular, we will work with the full $6\times 6$ covariance matrix
computed at an epoch in the middle of the observational arc of the asteroid.

\subsection{The Eclipse Plane}

\begin{figure}
\psfrag{s}{\rotatebox{13.7}{To the Sun}}
\psfrag{b}{Body}
\psfrag{a}{Asteroid}
\psfrag{ep}{Eclipse Plane}
\includegraphics*[width=\columnwidth]{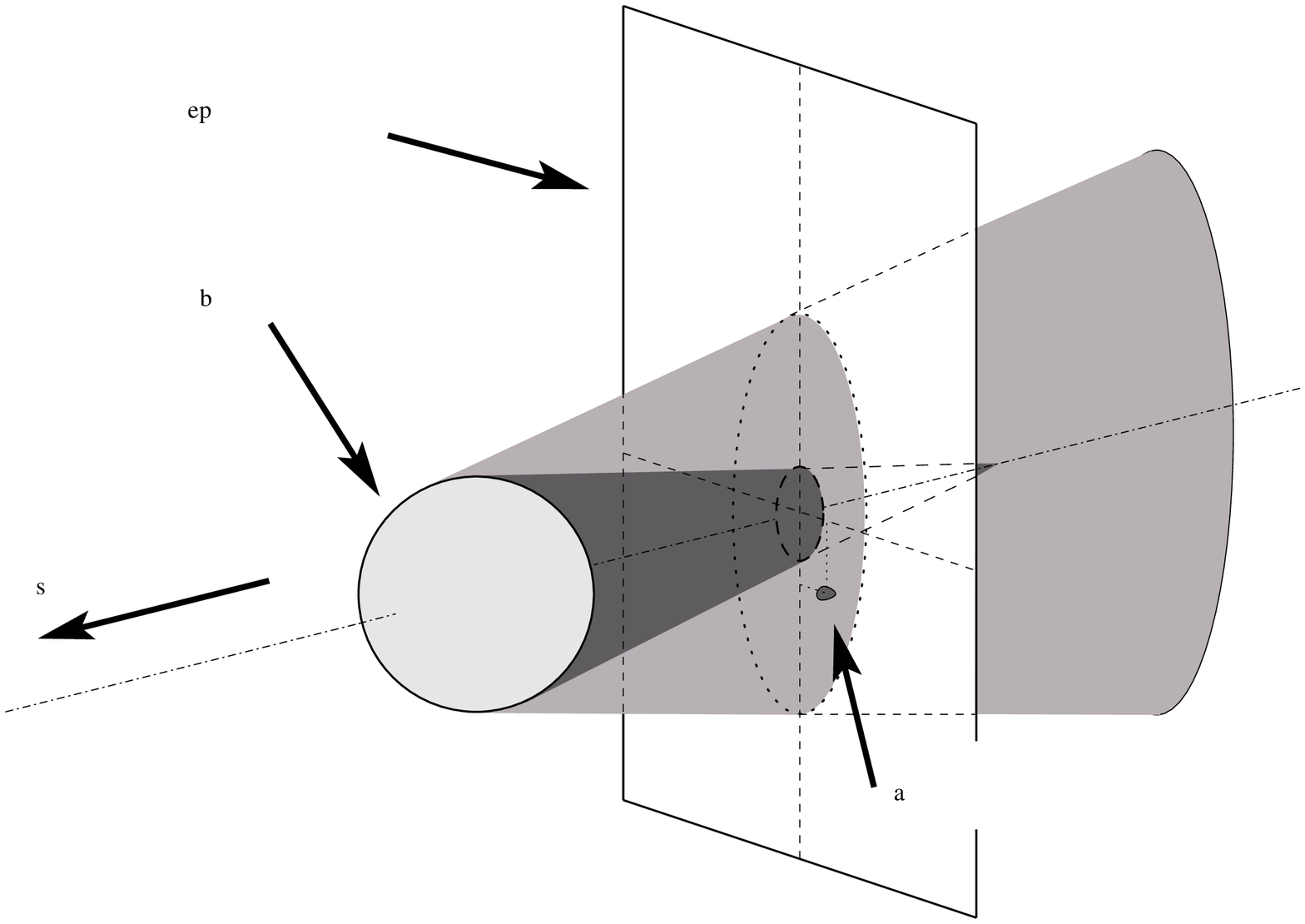}
\caption{
Eclipse Plane.
The light is coming from the Sun on the left side of the diagram, and
the penumbra and umbra cones are visible.
At any time, the eclipse plane is orthogonal to the Sun-body line,
and contains the asteroid.
\label{fig_ep}
}
\end{figure}

Given the positions of the Sun, the eclipsing body and an asteroid,
the EP is defined as the plane containing the asteroid,
orthogonal to the Sun-body line (see Fig.~\ref{fig_ep}).
As the asteroid's position changes with time,
the position of the EP is updated and a new data 
point is added to the plane.
When the asteroid is represented by a cloud of VAs
(see Section~\ref{va_section}),
each VA contributes data points to a different EP.
Every data point in the plane represents the VA at a different time,
and the three possible VA states are represented:
outside penumbra, inside penumbra, and inside umbra.
When all the different EPs are generated,
they're stacked, to obtain the final asteroid's EP.

The eclipse probability can be estimated simply as the fraction of VAs eclipsed,
with two different probabilities for penumbra and umbra eclipse.
The sensitivity of this probability estimate is clearly limited by the
number of VAs used.
For the purpose of this work, we are interested in probabilities of
the order of 0.01 or greater, achieved by using 512 VAs for each asteroid,
in order to limit fluctuations.

\subsection{Apparent Magnitude Computations \label{sect_mag_comp}}

During the eclipse, the apparent magnitude $V$ of an asteroid is
\begin{equation}
V = H + 5 \log_{10}(d \cdot r) + P(\phi,G) + Q(\gamma)
\label{eq_mag}
\end{equation}
with $H$ the absolute magnitude of the asteroid,
$d$ and $r$ the distances of the asteroid from the observer and the Sun
in AU, respectively,
$P(\phi,G)$ is a phase relation function,
$\phi$ is the phase angle (Sun-asteroid-observer angle),
$G$ is the slope parameter,
$Q(\gamma)$ eclipse correction,
and $\gamma$ is the ratio between the eclipsed Sun light flux and the nominal solar light flux.
The parameterization used for the phase relation function $P(\phi,G)$ is 
\begin{eqnarray}
P(\phi,G)    &=& -2.5 \log_{10}\left((1-G)\Phi_1(\phi)+G\Phi_2(\phi)\right) \nonumber \\
\Phi_1(\phi) &=& \exp\left(-3.33 (\tan(\phi/2))^{0.63}\right) \\
\Phi_2(\phi) &=& \exp\left(-1.87 (\tan(\phi/2))^{1.22}\right) \nonumber \ \ .
\end{eqnarray}
\citep{bowell_asteroids_2,limb_darkening}.
The value of $G$ adopted for all the asteroids is the default value $G =$~0.15.
The eclipse correction $Q(\gamma)$ is a simple function of $\gamma$:
\begin{equation}
Q(\gamma) = -2.5 \log_{10}(\gamma)
\end{equation}
and can be derived considering that,
if the light flux from a partially eclipsed solar disk is $\gamma F_{\sun}$
and the light flux from whole solar disk is $F_{\sun}$,
the apparent magnitude of the Sun, observed from the asteroid is 
$m_{\gamma} = -2.5 \log_{10}(\gamma F_{\sun} / F_{0}) = m_{\sun} -2.5 \log_{10}(\gamma)$, with $F_{0}$ the reference light flux for $m=0$.
The parameter $\gamma$ is in the range between 1 and 0,
with the limit values relative to the unobstructed Sun and to a total eclipse, respectively.
For the numerical computation of $\gamma$ we take into account the
eclipse geometry and the solar limb darkening.
The limb darkening effect \citep{limb_darkening} can be described by
\begin{equation}
\frac{I_\lambda(\psi)}{I_\lambda(0)} = 1 - u_2 - v_2 + u_2 \cos(\psi) + v_2 \cos^2(\psi)
\label{eq_sun_limb_intensity}
\end{equation}
with $\psi$ denoting the angle, measured from the Sun's center, between the observer and the point on the Sun's surface. 
Here, $I_\lambda(\psi)$ is the light intensity observed at wavelength $\lambda$.
The values of the constants $u_2$ and $v_2$ relative to 
$\lambda=550$~{nm}, used in the following computations,
are $u_2$~=~0.93 and $v_2$~=~-0.23 \citep{limb_darkening}.
The angle $\theta$ between the solar disk center and a point on the Sun's surface, as measured from the asteroid,
is related to $\psi$ through the transformation
\begin{equation}
\sin(\psi) = \sin(\theta) / \sin(\Omega)
\label{eq_angles}
\end{equation}
valid in the limit of $R \ll D$,
where $\Omega$ is the solar disk semi-aperture, and $0\le\theta\le\Omega$.

Using Eq.~(\ref{eq_sun_limb_intensity}) and Eq.~(\ref{eq_angles}),
it is possible to estimate numerically the fraction $\gamma$ of light
flux reaching the asteroid. In particular, using the normalization $N=1/(1-u_2/3-v_2/2)$
the integral of Eq.~(\ref{eq_sun_limb_intensity})
over the fraction $\Gamma$ of the solar disk visible from an asteroid
is always between 0 and 1.
%
The variable $\Gamma$ can be easily computed:
with $\Omega_s$ and $\Omega_b$ being the semi-aperture angles for the solar disk
and the eclipsing body disk, respectively, and $\zeta_{sb}$ the angle 
between the two disks centers, all measured from the asteroid,
we define $\beta_b$ and $\beta_s$ from
\begin{eqnarray}
\cos(\beta_b) &=& (\Omega_b^2+\zeta_{sb}^2-\Omega_s^2)/(2\Omega_b\zeta_{sb}) \\
\cos(\beta_s) &=& (\Omega_s^2+\zeta_{sb}^2-\Omega_b^2)/(2\Omega_s\zeta_{sb}) \ \ .
\end{eqnarray}
We then let
\begin{eqnarray}
A_b &=& \Omega_b^2 (\beta_b-\sin(\beta_b)\cos(\beta_b)) \\
A_s &=& \Omega_s^2 (\beta_s-\sin(\beta_s)\cos(\beta_s))
\end{eqnarray}
and the visible fraction $\Gamma$ of the solar disk,
sorted by decreasing value of $\zeta_{sb}$, is
\begin{equation}
\Gamma = \left\{
\begin{array}{ll}
1                                  & \Omega_s + \Omega_b  \le \zeta_{sb}  \\ 
1 - (A_b+A_s)/(\pi\Omega_s^2) \ \  & \Delta \le \zeta_{sb} < \Omega_s + \Omega_b \\
1 - \Omega_b^2/\Omega_s^2          & \zeta_{sb} < \Delta\textrm{, } \Omega_b < \Omega_s  \\
0                                  & \zeta_{sb} < \Delta\textrm{, } \Omega_s \le \Omega_b
\end{array}
\right.
\label{eq_big_gamma}
\end{equation}
with $\Delta = |\Omega_s - \Omega_b|$.
These four cases are relative to the body disk outside, crossing and inside the solar disk,
with the last case separated into two sub-cases,
based on the relation between $\Omega_s$ and $\Omega_b$.
In general, the difference between $\gamma$ and $\Gamma$
depends on the position of the body with respect to the solar disk.

\begin{figure}
\includegraphics*[width=\columnwidth]{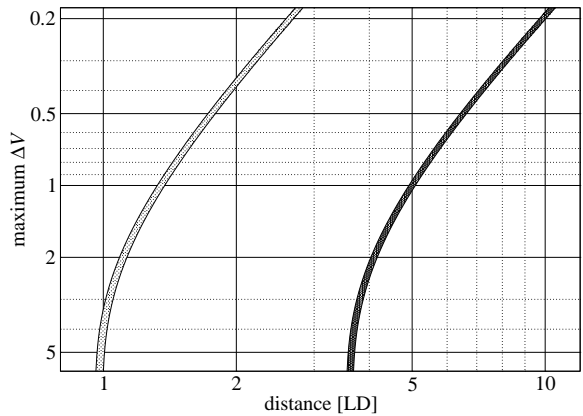}
\caption{
Maximum apparent magnitude effect due to an eclipse,
as a function of the distance between the asteroid and the Moon (left band,
light gray) or the Earth (right band, dark gray).
The width of each band is related to the range of the distance $D$
between the eclipsing body and the Sun.
\label{fig_max_dV}
}
\end{figure}

By using this formalism we can compute the maximum apparent magnitude effect
due to an eclipse for a generic asteroid,
as a function of the close approach distance from Earth or Moon,
keeping into account the Earth to Sun and Moon to Sun distance range.
The result is reported in  Fig.~\ref{fig_max_dV},
where the maximum effect is obtained by forcing the perfect
Sun-body-asteroid alignment, so that the body is obscuring the 
brighter part of the solar disk.


Two important classes of error have to be considered
when trying to predict an asteroid magnitude:
the light curve of the asteroid, 
and the atmosphere, oblateness and surface features of the eclipsing body.
An accurate light curve is required in order to
correctly interpret the observations.
The light curve amplitude of the asteroid
can be bigger than the eclipse effect,
and should be measured immediately before and after the eclipse,
in order to account for possible changes in the rotation due to the 
close approach to the eclipsing body, and also to account for
complex light-curves related to tumbling asteroids \citep{tumbling_asteroids}.

When the eclipsing body has an atmosphere, as is the case
for NEAs eclipsed by the Earth, the apparent magnitude V can differ
from that estimated here because of
refraction and scattering of sunlight through the atmosphere.
These effects become more and more important as a smaller fraction of the
solar disk is illuminating the asteroid. 
In particular, the asteroid's apparent magnitude $V$ will have a finite value
even during the total eclipse phase.
An estimate of this effect can be obtained from the Moon's apparent magnitude
during a total Moon eclipse: according to \cite{1993AAS...183.2703M},
the apparent magnitude of the Moon fades from -12.7 outside the eclipse to +1.4 
when the Moon reaches the maximum eclipse point.
This gives an estimate of the maximum effect of the eclipse,
$\Delta V_{\mathrm{max}} =$~14.1, for an object
at the distance of 1 LD from the Earth. This limit can be bigger by up to 5 magnitudes
after volcanic eruptions \citep{1993AAS...183.2703M}.
The maximum eclipse effect at distances different from 1 LD can
be obtained using inverse-square-law scaling.
Body oblateness and a non-smooth body surface can also 
represent sources of error to the magnitude estimate,
the former mainly during the penumbra eclipse,
the latter mainly at the umbra-penumbra interface.

\subsection{Eclipse Observation and Orbit Improvement\label{sec_orbit_improvement}}

An appreciable fraction of relatively small NEAs 
are discovered during their close approach to the
Earth-Moon system, and can be followed only for a
limited number of days.
This means that their observed arc is extremely
short at the moment of the close approach, and their orbit
can be determined with very limited accuracy.
In case of an eclipse during the passage of the asteroid, 
it is possible in principle to measure the main features of the eclipse:
the epoch relative to the beginning and the end of the total eclipse,
and the magnitude variations during the partial eclipse phase.
These data can help to improve the orbit accuracy:
intuitively, if a set of VAs is used to statistically describe
the orbital evolution of the asteroid, in general every VA will
enter and leave the umbra region at different times
and with a different geometry; the comparison
between observations and predictions for different VAs 
could restrict the space of parameters to a subset
having a behavior compatible with the observations.
For total eclipses, accurate measurements of the umbra entering and leaving
times are indirect observer-asteroid distance measurements:
the intersection between a 2D surface (the umbra cone) and
the line of sight from the observer to the asteroid 
provides the asteroid position within the measurement and model errors.

We propose two different methods to improve the asteroid orbit: 
one based on least squares minimization of a 
modified minimized function, and another based on VAs weighting.

The standard function minimized using least squares techniques is
the sum over all the astrometric observations
of the residuals $\delta$ on right ascension and declination
\begin{equation}
f = \sum_{\mathrm{obs}} \left( \left(\frac{\delta_{\mathrm{RA}}}{\sigma_{\mathrm{RA}}}\right)^2 + \left(\frac{\delta_{\mathrm{DEC}}}{\sigma_{\mathrm{DEC}}}\right)^2 \right) \nonumber
\end{equation}
with an estimated accuracy $\sigma$ of the order of 1.0 arcseconds for observations by modern telescopes using CCD cameras.
This function can be extended, to include the eclipse magnitude observations 
\begin{equation}
f' = f + \sum_{\mathrm{obs}} \left(\frac{\delta_{
\mathrm{mag}}}{\sigma_{\mathrm{mag}}}\right)^2 \nonumber
\end{equation}
with $\delta_{\mathrm{mag}}$ the magnitude residual
and $\sigma_{\mathrm{mag}}$ the estimate of the total error on the magnitude measurement.
A major problem related to this method 
is the correct estimate of $\sigma_{\mathrm{mag}}$,
taking into account the asteroid light curve 
and effect of the eclipsing body's atmosphere, oblateness and surface features.
We are still investigating the applicability of this method;
a direct asteroid eclipse observation would provide 
a great opportunity to test and validate this method.

An alternative approach is the following:
we select a subset of all the VAs by assigning to each VA a weight $w$ defined as
\begin{equation}
w^2 = \frac{1}{N_{\mathrm{obs}}} \sum_{\mathrm{obs}} \left(\frac{\delta_{\mathrm{mag}}}{\sigma_{\mathrm{mag}}}\right)^2 \nonumber
\end{equation}
and selecting all the VAs below a threshold $w_{\mathrm{max}}$.
In this case $\delta_{\mathrm{mag}}$ represents the magnitude residual
for the VA considered.
This method doesn't improve the nominal solution, but can be useful to
determine the possible sky projection of the selected VAs
after an observed eclipse, for recovery purposes.
As an example, see Section~\ref{sec_st26}
and Fig.~\ref{eclipse_mag_st26_earth_distance} for the case of 2004~ST$_{26}$.

\section{NEAs Eclipsed by Earth and Moon\label{sec_application}}

As an application of this eclipse prediction method,
we investigate the possibility for each known NEA
to be eclipsed by the Earth or the Moon.
We have propagated numerically the orbit of all the known NEAs,
3444 asteroids, in the period 1990-2050,
using the orbits and covariance matrices computed by the 
NEODyS group \citep{neodys}
that were available on August 17, 2005.
The dynamical system
includes the Sun, all planets, and the Moon, plus the massless VAs.
The numerical integration is performed using the 
ORSA\footnote{\tt http://orsa.sourceforge.net} framework.
In particular, a modified version of the  15th order RADAU integrator \citep{radau}
has been used, with a nominal accuracy of $10^{-12}$.
The difference between the original algorithm and the modified version
is that this modified version doesn't integrate the Solar 
System objects, as their positions and velocities 
are read at each internal timestep from the JPL ephemeris DE405
to ensure the maximum possible accuracy, 
while the VAs are numerically propagated.
The accuracy of this modified RADAU integrator,
as implemented in the ORSA framework,
has been extensively tested during the development of ORSA.
In particular, the minimum Earth/asteroid distance is always in
very good agreement (i.e.~well within the orbital elements uncertainties) with the data provided by the NEODyS website \citep{neodys}.

The results of this analysis have been collected in 
Table~\ref{asteroid_eclipse_table}. 
In particular, we have recorded all the eclipses with a 
penumbra eclipse probability $P_p \ge$~0.01
and a fraction of solar disk visible $\Gamma \le$~0.99,
that is, all the eclipses within 10~LD if by the Moon,
and within 40~LD if by the Earth.
A total of 74 distinct eclipses have been found, involving 59 asteroids.
Of the 74 eclipses found, 58 are by Earth and 16 by the Moon.
A detailed analysis of the probability of an asteroid eclipse
and an estimate of the expected number of eclipses per year
are provided in Section~\ref{sec_prob}.

We have found three classes of \emph{special} asteroid eclipses:
\emph{double eclipses}, \emph{multiple eclipses}, and \emph{immediate eclipses}.
These classes are not mutually exclusive, and we have examples of eclipses
belonging to more than one class.
An asteroid experiences a \emph{double eclipse}
if it is eclipsed by both Earth and Moon
during the same close approach;
more details are provided in Section~\ref{sec_double_ecl}.
A different pattern is observed in 3 other cases, when the same asteroid 
experiences several eclipses during different close approaches.
We will refer to these events as \emph{multiple eclipses}
(see Section~\ref{sec_multiple_ecl}).
Another class is represented by
asteroids with an eclipse happening during the same close approach as the discovery;
we will refer to these events as \emph{immediate eclipses},
and the asteroids with eclipses belonging to this class are listed in Table~\ref{immediate_table}
and studied in more detail in Section~\ref{sec_imm_ecl}.

In the following sections we describe some particular cases found,
beginning with the extremely unusual case of \apo.

\subsection{\apo}

The asteroid \apo{} was discovered on 2004 June 19 
by Roy Tucker, David Tholen, and Fabrizio Bernardi
while observing from the Kitt Peak National Observatory, Arizona.
Announced by \cite{2004MPEC....Y...25G},
this asteroid has set the highest level of attention since the introduction
of the \emph{Torino Scale} \cite[TS,][]{torino_scale},
reaching the TS-4 level.
Pre-discovery observations of \apo{} excluded the possibility of an impact with
the Earth on 2029 April 13, but still leaving the possibility 
of an impact at successive epochs.
Radar observations in January 2005 \citep{radar_iauc}
and September 2005 \citep{2005IAUC.8593....1G}
further improved the orbit of the asteroid.
In particular, for our computations we are using the orbit solution
computed including all the radar observations mentioned above.

\begin{figure}
\includegraphics*[width=\columnwidth]{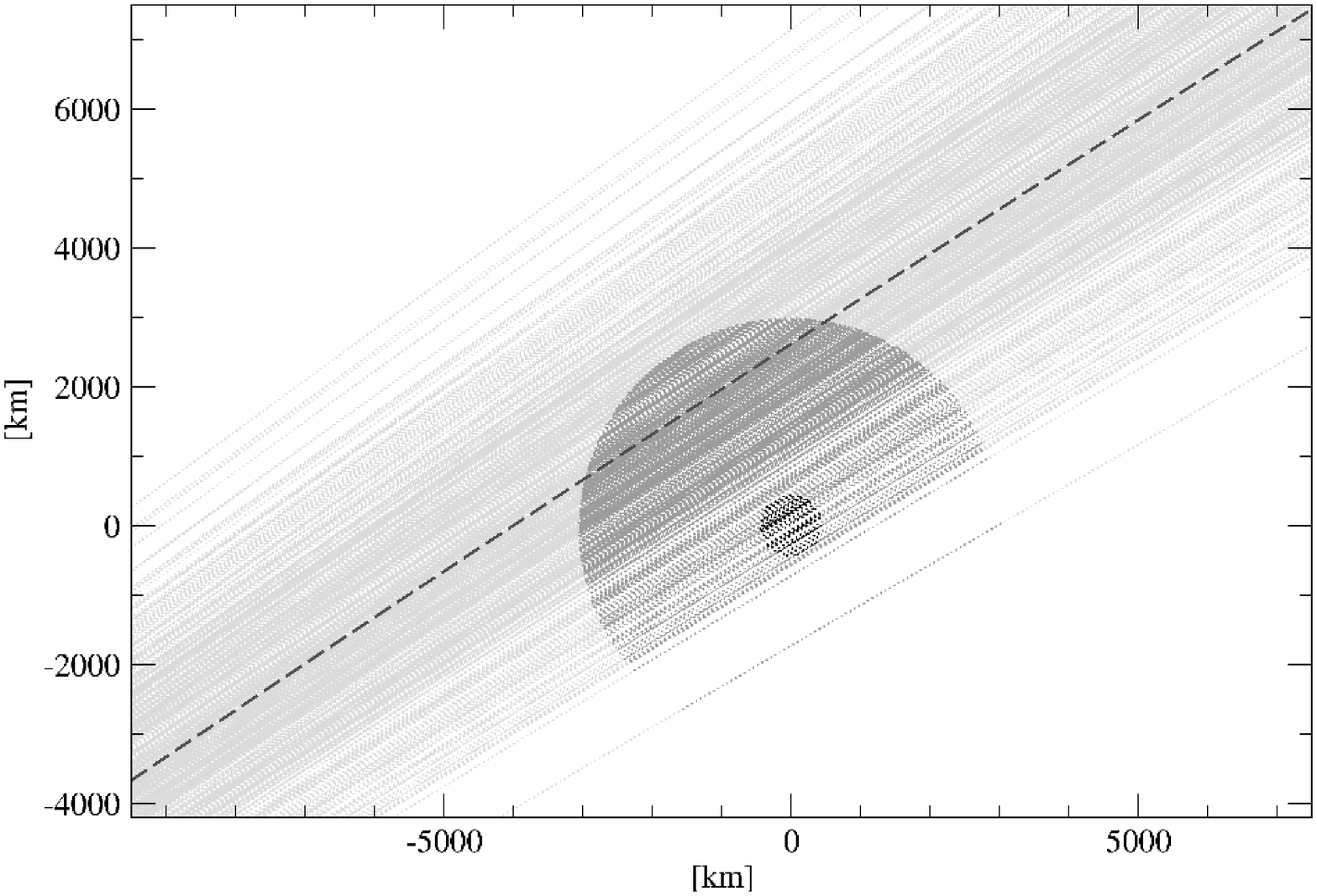}
\caption{
Eclipse Plane for \apo, eclipsed by the Moon on 2029 Apr 14 around 3:10 TDT.
The dark gray circular area represents the penumbra region, 
while the black circular spot is the umbra region.
The $x$ axis is parallel to the Ecliptic plane.
The light gray cloud of VAs is moving from bottom left to top right, and is slightly expanding
because it just experienced the close encounter with the Earth.
The nominal orbit, represented by the dashed line, enters the penumbra
region only.
About 74\% of all the VAs enter the penumbra region,
and 6\% of all the VAs enter the umbra region.
\label{eclipse_mn4}
}
\end{figure}

\begin{figure}
\includegraphics*[width=\columnwidth]{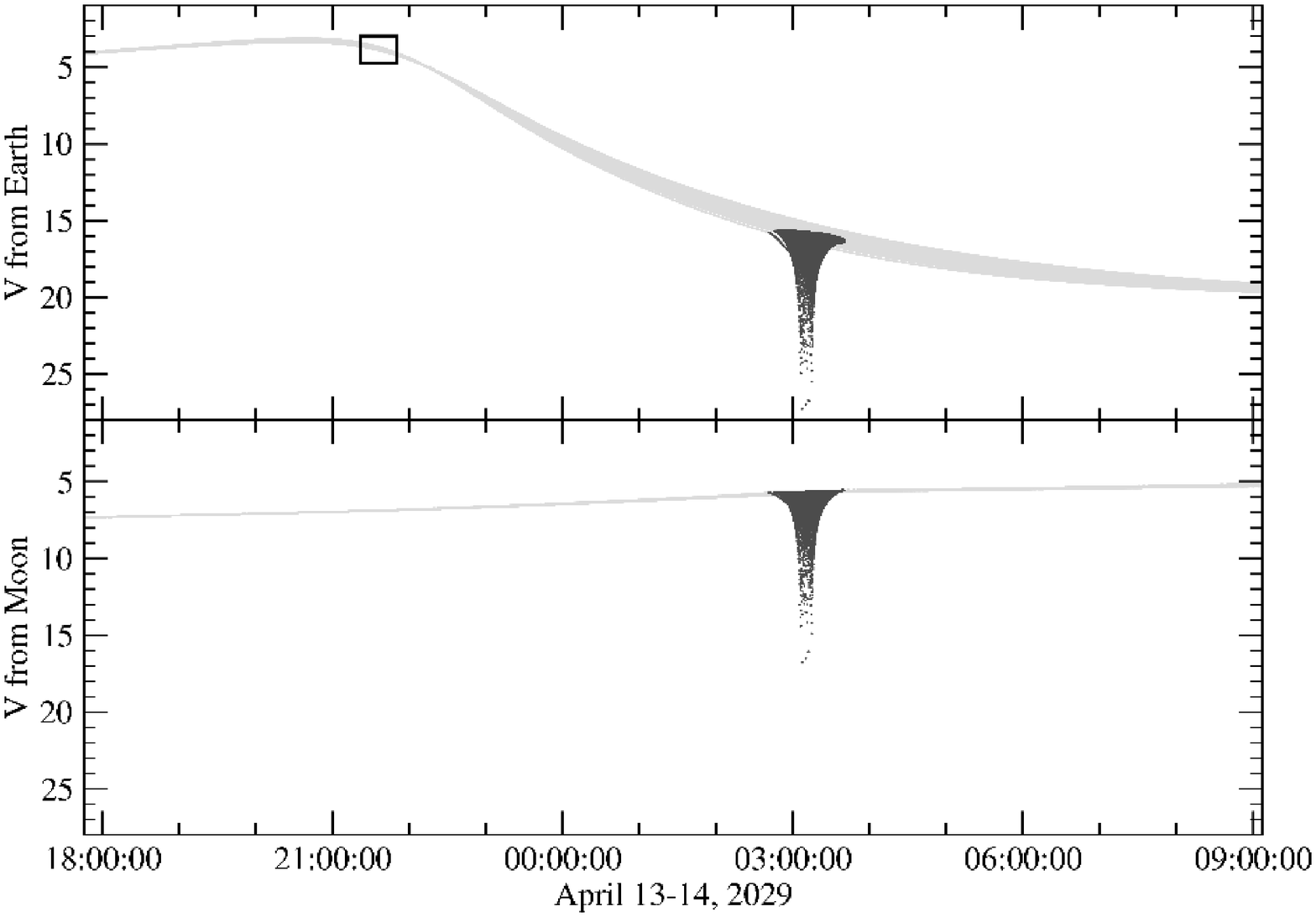}
\caption{Magnitude curve for \apo.
The upper half of the figure represents the apparent magnitude V of the
asteroid as observed from Earth, while the lower half represents the apparent magnitude V
as observed from the Moon.
The peaks around time 3:10 are the effect of the eclipse.
The small box near the top left corner of the plot 
represents the point of closest approach to Earth; its offset with respect to the 
brightness maximum
is due to the effect of the phase angle $P(\phi,G)$ in the magnitude formula, see Eq.~(\ref{eq_mag}).
The brightness peak is located about one hour earlier than the close approach time.}
\label{eclipse_mag_mn4}
\end{figure}

Less than one day after the close approach to Earth on 2029 April 13,
this asteroid has a chance of being eclipsed by the Moon.
The passage of the VAs close to or inside the Moon's shadow is
visible in Fig.~\ref{eclipse_mn4}.
The VAs cross the Moon's shadow cones at a distance between
0.70 and 0.75 LD from the Moon,
where the penumbra disc radius ranges between 
2981 and 3081 km (increasing with distance), 
and the umbra disc between 395 and 495~km (decreasing with distance).
About 74\% of all the VAs enter the penumbra,
starting as early as 2029 Apr 14 at 2:41 TDT (TDT is the Terrestrial Dynamical Time timescale)
and leaving it no later than 3:41 TDT of the same day, lasting up to 42 minutes, 32 minutes on average.
A smaller fraction, about 6\% of all the VAs,
enter the umbra starting as early as 3:06 TDT and
leave it no later than 3:14 TDT, 
lasting up to about 6 minutes, with an average of 4 minutes.
The VA relative to the nominal orbit enters the penumbra at 3:12 TDT 
and leaves it at 3:40 TDT
without entering the umbra (see Fig.~\ref{eclipse_mn4}).
%
This eclipse will not be visible from Earth because the solar elongation of this asteroid will range between about 15.9$^{\circ}$ and 17.7$^{\circ}$ during the eclipse.
A diagram showing the expected magnitude curve for \apo{}
is shown in Fig.~\ref{eclipse_mag_mn4}.
Simulated magnitudes are shown for observations both from Earth (or from a low orbit satellite) and from the Moon (in case a telescope will be in place 
on the side of the Moon facing the Earth by 2029).

\subsection{2004~ST$_{26}$ \label{sec_st26}}

\begin{figure}
\includegraphics*[width=\columnwidth]{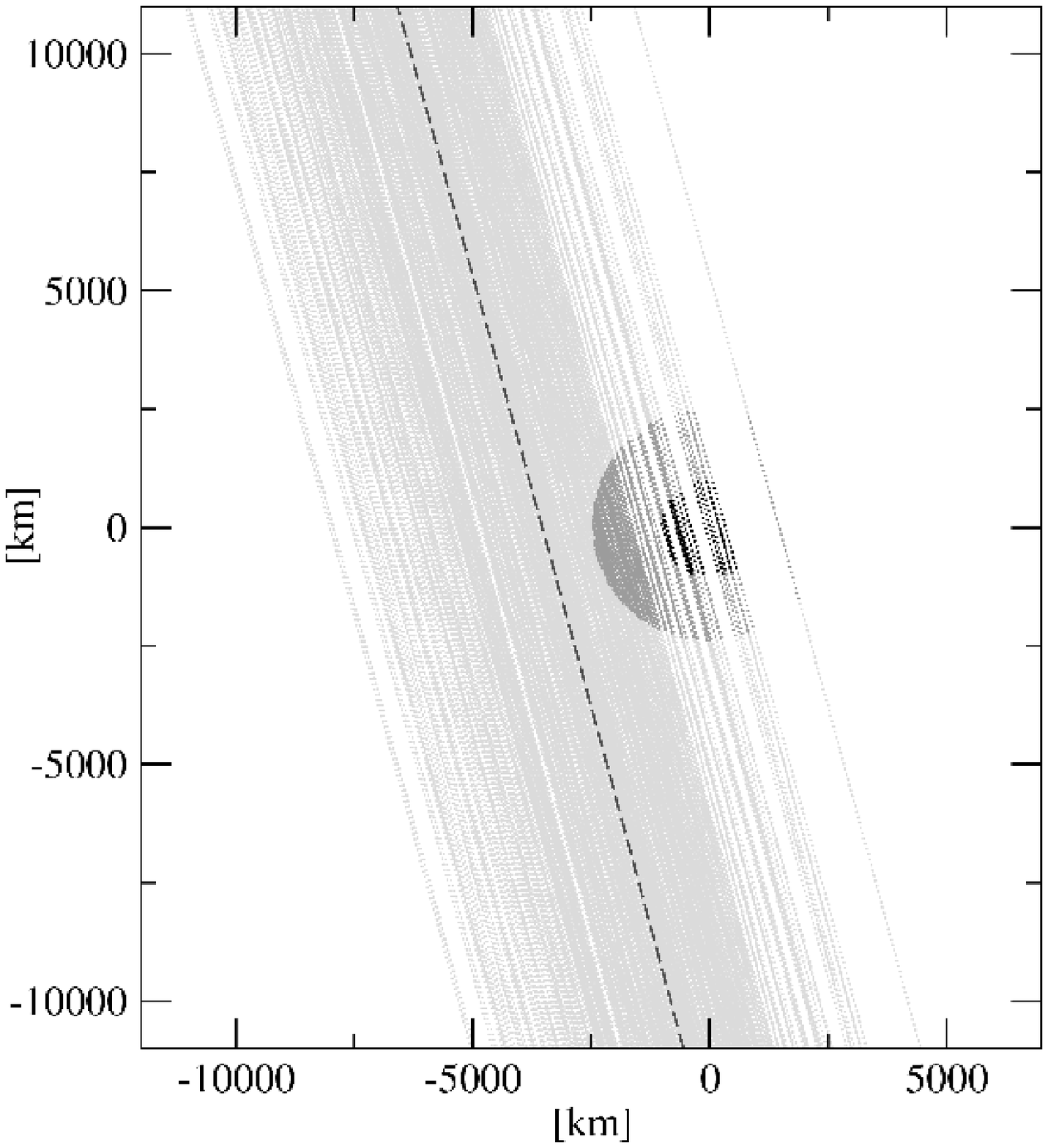}
\caption{Eclipse plane for 2004~ST$_{26}$,
eclipsed by the Moon on 2004 September 22 around 2:08 TDT.
About 18\% of all the VAs enter the penumbra region,
and 4\% of all the VAs enter the umbra region.
In this case, the VA relative to the nominal orbit,
represented by the dashed line,
never crosses the penumbra region.}
\label{eclipse_st26}
\end{figure}

This asteroid has a 18\% probability that it crossed the lunar penumbra
on 2004 September 22 at 2:08 TDT,
and about 4\% probability of having crossed the umbra cone.
The eclipse plane is provided by Fig.~\ref{eclipse_st26}.
%

\begin{figure}
\includegraphics*[width=\columnwidth]{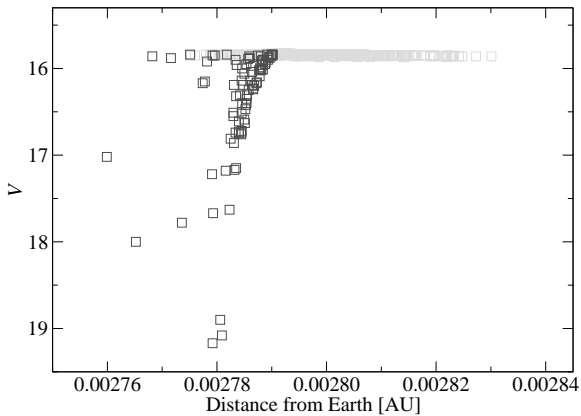}
\caption{Estimated magnitude distribution for all the VAs relative to 2004~ST$_{26}$,
as a function of the distance from Earth, at 2:08 TDT on September 22, 2004.
Light gray and dark gray symbols are relative to VAs outside and inside the
penumbra cone, respectively.}
\label{eclipse_mag_st26_earth_distance}
\end{figure}

This case can be used to show the ideas behind the orbit improvement connected 
with the direct observation of an asteroid eclipse, as described in Section~\ref{sec_orbit_improvement}.
In Fig.~\ref{eclipse_mag_st26_earth_distance} we have the distribution of apparent magnitude,
as a function of the distance from Earth, for all the generated VAs at a fixed time.
Each data point represents a different VA.
An eclipse observation at this time could select, 
within the magnitude measurement errors,
the fraction of VAs compatible with observations:
As an example, an apparent magnitude measurement of $V=19.0 \pm 1.0$
would indicate a distance from Earth of about 0.00278~AU,
promoting VAs (and relative orbits) at that distance.
Additional information on the apparent
magnitude curve, obtained with successive measurements,
could further improve the relative weight of different asteroids
to help in the selection of the most probable region in the space
of the orbital elements.
%

\subsection{2003~WT$_{153}$}

\begin{figure}
\includegraphics*[width=\columnwidth]{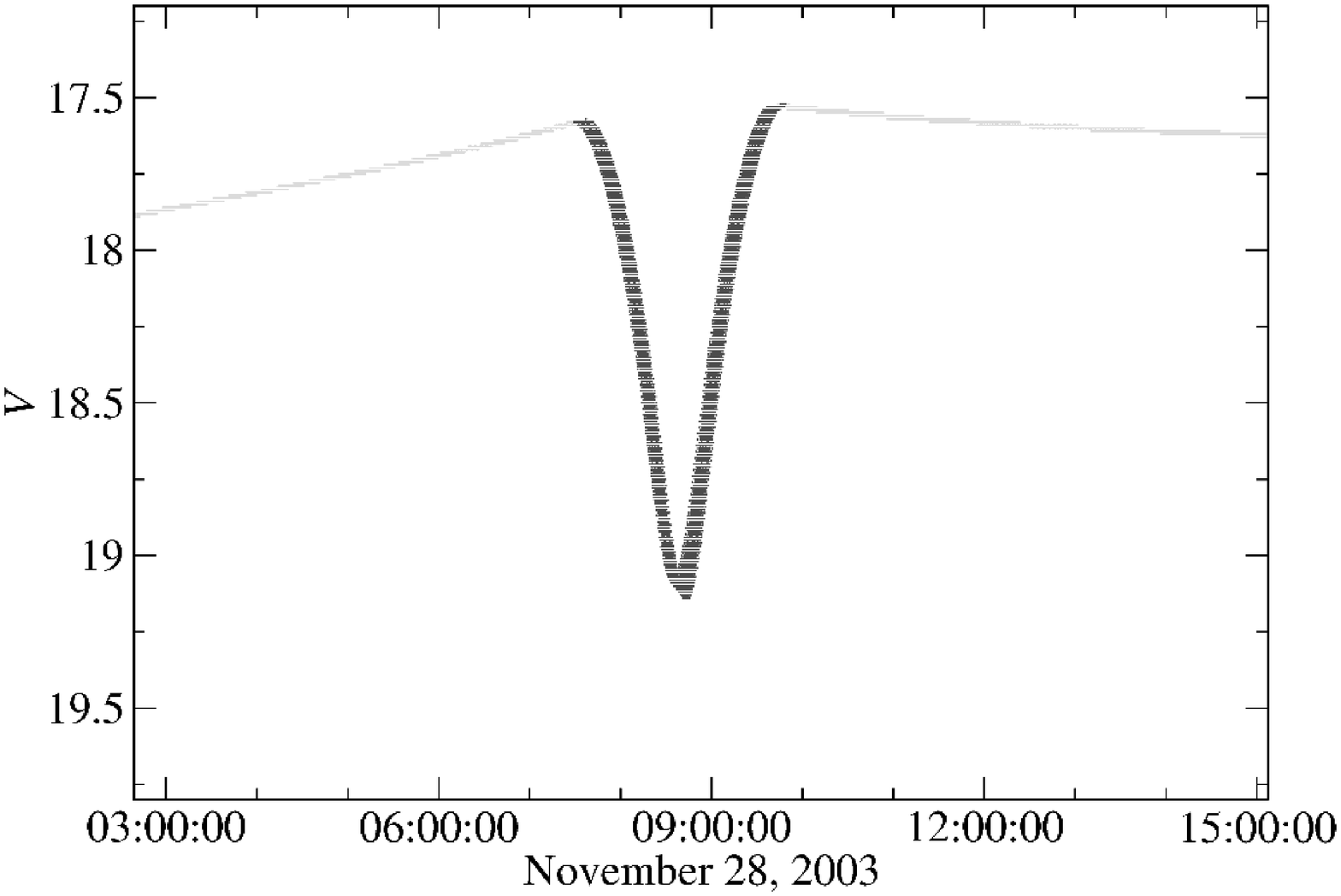}
\caption{Magnitude curve for 2003~WT$_{153}$,
eclipsed by the Earth on 2003 Nov 28 at 8:44 TDT.
All the VAs relative to this asteroid crossed the Earth's penumbra cone.}
\label{eclipse_mag_wt153}
\end{figure}

The asteroid 2003~WT$_{153}$ experiences one of the 15 eclipses 
found with penumbra eclipse probability $P_{p} = 1.00$ (see Table~\ref{asteroid_eclipse_table}),
that is, all the VAs relative to this asteroid crossed the Earth's penumbra
cone on 2003 Nov 28 at 8:44 TDT.
This eclipse, not directly observed, lasted about 130 minutes,
with an important effect on the apparent magnitude
(see Fig.~\ref{eclipse_mag_wt153}).


\subsection{2003~SY$_{4}$ \label{sec_SY}}

This asteroid is the only one found, a posteriori, with astrometry measurements during an eclipse.
In particular, part of the discovery observations were made during the eclipse
listed on Table~\ref{asteroid_eclipse_table}.
Discovered by M.~Block at the LPL/Spacewatch~II observatory \citep{2003MPEC....S...12T}
on 2003 September 17 with a series of six observations, it turns out that the
first two observations, at 6:58 TDT and 7:19 TDT, were inside
the eclipse window. The asteroid 2003~SY$_{4}$ has been eclipsed 
by the Earth on 2003 September 17 between 6:48 TDT and 7:24 TDT, 
and it crossed, with a probability of 100\%, the penumbra cone.
Since the minimum solar disk fraction visible during the eclipse is 0.98,
the effect of the eclipse is much smaller than the magnitude measurements error:
the measured apparent magnitude during the eclipse is $V_{\mathrm{obs}}=17.9$
while the computed one is $V_{\mathrm{comp}}=17.7$ 
(neglecting all the 
error sources described in Section~\ref{sect_mag_comp}).

\subsection{Double Eclipses\label{sec_double_ecl}}

We have found 9 different double eclipses, 
i.e., eclipses when the same asteroid is eclipsed by both Earth and the Moon during the same close approach, usually within the span of a few hours.
The 9 events found in our numerical simulations are:
2004~TB$_{10}$  in 1990,
2003~SW$_{130}$ in 1990,
1998~BT$_{13}$  in 1998,
2001~FE$_{90}$  in 2009,
2004~BL$_{86}$  in 2015,
2001~GP$_{2}$   in 2020,
2004~UT$_{1}$   in 2022,
1999~VX$_{25}$  in 2040,
2005~KA         in 2042.

The case of 1998~BT$_{13}$ is the only one with  probability 1.00 for both penumbra eclipses.
The case of 2004~UT$_{1}$ is also interesting
because the two eclipses can be \emph{connected}, i.e.~the asteroid can be eclipsed
by Earth and Moon at the same time, because the second eclipse by the Earth
can start before the first eclipse, by the Moon, is finished.

\subsection{Multiple Eclipses\label{sec_multiple_ecl}}

We have \emph{multiple eclipses} when the same asteroid 
experiences several eclipses during different close approaches.
We have found 3 such cases:
2001~LB         (2012, 2023),
2003~WT$_{153}$ (2003, 2030),
2003~YT$_{70}$  (1999, 2001, 2005, 2007, 2009).
In particular, all the multiple eclipses found involve the Earth.
The multiple eclipses pattern for the asteroid 2003~YT$_{70}$ is related 
to its orbital period being very close to 2 years.

\subsection{Immediate Eclipses\label{sec_imm_ecl}}

We have \emph{immediate eclipses} every time an asteroid experiences an eclipse 
during the same close approach as its discovery.
We have collected all 18 eclipses belonging to this class in Table~\ref{immediate_table}.
In particular, 11 immediate eclipses are by the Earth and 7 by the Moon.
We have investigated the relation between the absolute magnitude $H$ of the asteroid involved
and the delay $\Delta t = t_{e} - t_{d}$ between eclipse time $t_{e}$ and 
asteroid discovery time $t_{d}$.
%
\begin{figure}
\includegraphics*[width=\columnwidth]{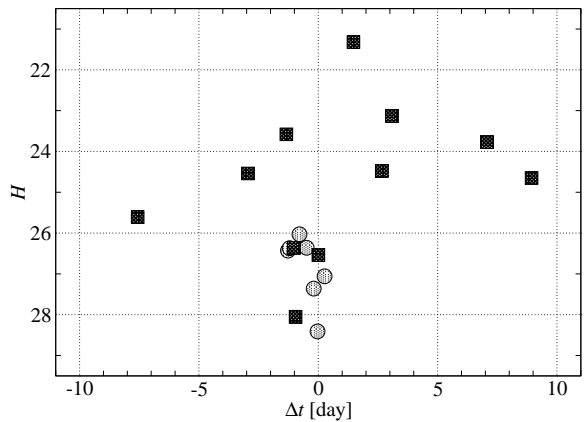}
\caption{
Immediate eclipses: asteroid absolute magnitude $H$
versus delay $\Delta t$ (see text).
The square symbol represents eclipses by the Earth,
while the circle represents eclipses by the Moon.
\label{fig_mag_delayeclipse}
}
\end{figure}
As Fig.~\ref{fig_mag_delayeclipse} suggests, we can distinguish between
immediate eclipses by bright asteroids ($H\lesssim 26$)
or by dim asteroids ($H\gtrsim 26$).
The 8 eclipses of bright asteroids (upper half of Fig.~\ref{fig_mag_delayeclipse},
all by Earth) exhibit delays of several days, 
with a tendency toward eclipse after discovery,
while the 10 eclipses of dim asteroids (lower half of Fig.~\ref{fig_mag_delayeclipse},
3 by Earth, 7 by Moon)
have much shorter delays,
with a strong tendency toward eclipse before discovery.

To try to explain this behavior we must consider that dim asteroids,
with an estimated diameter typically between 10 and 30 meters,
can be observed by many active telescopes only for a limited amount of time (often only 2 or 3 nights),
and for this reason dim asteroids can be discovered only when the observer/asteroid distance is close to the minimum.
Also, even if potentially visible, a dim asteroid can remain undetected for several nights
because of many factors related to the observatory hardware,
asteroid dynamics, and other selection effects (see \cite{2002aste.conf...71J}
for a review).
Now, if we consider that an eclipse by Earth or the Moon can take place only at a
time close to the middle of the discovery time window,
and that this discovery window is small for dim asteroids and relatively large for bright ones,
it is easy to realize that
we have only 1 or 2 nights to discover a dim asteroid before the eclipse,
while we have several nights to discover a bright asteroid before the eclipse.
Since 2 nights can often be insufficient to discover a dim asteroid,
the eclipse is likely to happen before the dim asteroid is discovered.
This effect poses substantial limits to the numerical prediction and the direct observation of immediate eclipses
involving dim asteroids.

\section{Eclipse Probability for NEAs\label{sec_prob}}

We attempt now to estimate the probability of two different events:
the average eclipse probability per NEA,
and the probability for an eclipse to occur at an epoch very close to the discovery epoch of the asteroid.

The average eclipse probability per NEA can be obtained
by properly scaling the impact probability between a NEA and the Earth, that
\cite{stuart_binzel_2004} estimate as $1.50 \times 10^{-9}$ per year per NEA.
The scaling factor is the ratio between the eclipse cone longitudinal section
and the gravitational capture radius disk of the Earth. This section
is well approximated by a trapezoid for the penumbra cone, 
and by a triangle for the umbra cone.
By using the relations developed in Section~\ref{sec_eclipse_geom},
and using the Earth's gravitational capture radius equal to 7540~km \citep{stuart_binzel_2004}
when dealing with impact probabilities and the Earth's equatorial radius of 6378~km
for geometric computations, and limiting the penumbra cone length to 10~LD for the Moon and to 40~LD for the Earth
(that is, about ten times the respective umbra cone length),
with a simple algebra we find that the penumbra and umbra scaling factors for the Earth are, respectively,
$\sigma_{p,E} \simeq$~7300 and $\sigma_{u,E} \simeq$~50,
while for the Moon we get $\sigma_{p,M} \simeq$~460 and $\sigma_{u,M} \simeq$~4.
The average eclipse probabilities are 
$P_{p,E} \simeq 1.1 \times 10^{-5}$,
$P_{u,E} \simeq 7.5 \times 10^{-8}$,
$P_{p,M} \simeq 6.9 \times 10^{-7}$,
$P_{u,M} \simeq 6.0 \times 10^{-9}$,
all per year per NEA.

We can scale these probabilities to the number of NEA (3444 asteroids) examined in our analysis,
and to the 60 years period monitored, obtaining
the expected number of events
$N_{p,E} \simeq 2.26$,
$N_{u,E} \simeq 0.015$,
$N_{p,M} \simeq 0.14$ and
$N_{u,M} \simeq 0.0012$.
These probabilities can be compared directly to the probabilities of the eclipses found (see Table~\ref{asteroid_eclipse_table}),
considering the eclipse probability as a \emph{fractional event}.
The sum $S$ of all the probabilities relative to asteroids with eclipses at epochs far from the discovery epoch,
i.e.~considering only the asteroids that are in Table~\ref{asteroid_eclipse_table} but not in Table~\ref{immediate_table},
gives $S_{p,E} = 3.40$,
$S_{u,E} = 0.02$, 
$S_{p,M} = 0.93$ and
$S_{u,M} = 0.06$.
The agreement between estimated and numerically detected number of events
for eclipses by the Earth can be considered satisfactory,
considering the relatively small number of eclipses found. 
On the other hand, the sums $S_{p,M}$ and
$S_{u,M}$ for eclipses by the Moon
are significantly different from the estimates $N_{p,M}$ and
$N_{u,M}$, with \apo{} as the major contributor.
We notice that, neglecting the contribution from \apo{}, 
we would have $S_{p,M} = 0.19$ and
$S_{u,M} = 0.00$, in satisfactory agreement with the predicted probability.
To partially justify this solution to the discrepancy we notice that \apo's
orbit has been determined with an accuracy that is orders of magnitude
better than the average accuracy of all the other asteroids experiencing an eclipse,
putting this asteroid somewhat out of statistics.
An alternative explanation, still under investigation,
is of dynamical nature: the Moon's penumbra and umbra cones can intersect
a number of NEAs orbits larger than expected because the Moon, while orbiting
around the Earth,
is allowed in general to get a bit closer to an asteroid,
especially when compared to an hypothetic Moon fixed on the Earth's orbit.

What is the probability for a newly discovered NEA to experience an eclipse during the 
same close approach of the discovery?
We can estimate it using the NEA discovery statistics \citep{jpl_neo_stats}.
Our analysis extends from 1990 to 2050, but for this section we are limited 
to the period between 1990 and today.
The NEAs discovered between Jan 1, 1990 and August 1, 2005 are $3434-134=3300$, and in this
period only 17 asteroids have experienced eclipses at a time very close to the discovery time
(for a total of 18 eclipses, because 1998~BT$_{13}$ experiences a double eclipse).
This leads to a probability of about one in 194 for a newly discovered NEA
to experience an eclipse at a time very close to the discovery time.
If we limit ourselves to the period between Jan 1, 1998 and August 1, 2005
only $3434-445=2989$ NEAs have been discovered in this period,
and only 16 of the original 17 asteroids fall into this period.
The probability estimate for this restricted period is of about one in 187.
With 439 NEAs discovered in 2003 and 532 discovered in 2004 \citep{jpl_neo_stats},
a number of events between two and three per year can be expected,
and possibly more, as the six events in 2003.

\section{Conclusions}

Near Earth Asteroids can be eclipsed by Earth and the Moon as has happened in the past 15 years, and will occur again in the
future to known asteroids and to asteroids not yet discovered.
We have compiled a catalog, based on numerical simulations,
of all the asteroid eclipses between a NEA and Earth or the Moon,
in the period 1990-2050, including a total of 74 distinct eclipses involving 59 asteroids.
A closer inspection at these eclipses allowed us to determine
three different recurring patterns: 
\emph{double eclipses}, \emph{multiple eclipses}, and \emph{immediate eclipses}.
In particular, the analysis of \emph{immediate eclipses} allowed
us to explain their apparent tendency to happen within one or two days before
the asteroid discovery.

An analysis of the NEAs eclipse probability offers two main results:
$i)$ asteroid impact probabilities for Earth and the Moon can be scaled
in order to predict asteroid eclipse probability, with satisfactory
agreement between predicted and numerically detected number of events;
$ii)$ every newly discovered NEA has a probability of about one in 190 
to experience an eclipse during the same close approach of the discovery.
Given the present rate of discovery, several new eclipses are expected every year involving newly discovered asteroids.
With the beginning of a new class of NEOs surveys, 
such as LSST \citep{lsst_2004} and Pan-STARRS \citep{pan_starrs_2005},
it will be possible to cover the entire visible sky to the 24th magnitude in less than a week,
enabling discovery rates almost two orders of magnitude greater than all existing surveys combined,
hence allowing the direct observation of up to a few hundred NEA eclipses per year.
We plan to routinely monitor numerically all NEAs for potential eclipses, using the
\emph{Distributed Computing System for Near Earth Objects Hazard Monitoring}
\citep{boinc_orsa} now under construction.

The direct observation of an asteroid eclipse can provide data useful to improve the orbit
of the asteroid. We suggest two different methods to improve the asteroid orbit, 
the first as an extension of the standard least squares method, and the second as a weighting rule for different VAs relative to the asteroid.
We plan to test and extend these methods as soon as data relative to direct observations 
of NEA eclipses will be available.
%

The theoretical tools developed in this work in order to statistically 
predict an asteroid eclipse
apply to any case of an asteroid eclipsed by a larger spherical body.

\acknowledgments


We gratefully acknowledge
Don Davis,
Beatrice Mueller,
Steve Kortenkamp
and 
Stu Weidenschilling
for their suggestions,
comments,
and constructive criticisms.



{
\small

}

\include{asteroid_eclipse_table}

\include{immediate_table}

\end{document}

%% file: asteroid_eclipse_table.tex
\begin{deluxetable}{lllrrrrcrccrrrc} 
\rotate
\tablecolumns{15}
\tabletypesize{\scriptsize}
\tablewidth{0pc}
\tablecaption{Asteroids Eclipsed by Earth and Moon.\label{asteroid_eclipse_table}}
\tablehead{
\colhead{} &  
\colhead{} &  
\colhead{} &  
\multicolumn{4}{c}{\emph{penumbra}} &
\colhead{} &  
\multicolumn{3}{c}{\emph{umbra}} 
\\
\cline{4-7} \cline{9-11} 
\\
\colhead{asteroid} & 
\colhead{body} &
\colhead{epoch} &
\colhead{$P_{p}$} &
\colhead{$\langle T_{p} \rangle$} &
\colhead{max($T_{p}$)} &
\colhead{min($\Gamma$)} &
\colhead{} & 
\colhead{$P_{u}$} & 
\colhead{$\langle T_{u} \rangle$} &
\colhead{max($T_{u}$)} &
\colhead{$d_E$} &
\colhead{$\lambda$} &
\colhead{H} &
\colhead{diameter}
\\
\colhead{} & 
\colhead{} & 
\colhead{[TDT]} &
\colhead{} &
\colhead{[s]} &
\colhead{[s]} &
\colhead{} &
\colhead{} & 
\colhead{} & 
\colhead{[s]} &
\colhead{[s]} &
\colhead{[LD]} &
\colhead{[$^{\circ}$]} &
\colhead{} & 
\colhead{[m]}
}
\startdata
2004~TB$_{10}$    & Earth      & 1990 Apr  2.851 & 0.01 & 1900 &  2500 & 0.40 & & 0.00 & $-$ & $-$ & 4.7  & 179.7 & 21.25 &            160 $-$  370 \\
2004~TB$_{10}$    & Moon       & 1990 Apr  2.985 & 0.01 & 1200 &  1400 & 0.95 & & 0.00 & $-$ & $-$ & 4.7  & 168.0 & 21.25 &            160 $-$  370 \\
2003~SW$_{130}$   & Earth      & 1990 Sep 20.531 & 0.06 & 4600 &  5900 & 0.00 & & 0.01 & 300 & 400 & 3.1  & 179.6 & 29.12 &   \phantom{0}1 $-$   10 \\
2003~SW$_{130}$   & Moon       & 1990 Sep 20.870 & 0.05 & 4800 &  6100 & 0.95 & & 0.00 & $-$ & $-$ & 3.7  & 174.0 & 29.12 &   \phantom{0}1 $-$   10 \\
2000~CK$_{59}$    & Earth      & 1991 Feb  8.776 & 0.01 &18200 & 20000 & 0.95 & & 0.00 & $-$ & $-$ & 17.0 & 179.8 & 23.97 &  \phantom{1}40 $-$  100 \\
1991~TU           & Moon       & 1991 Oct  7.279 & 0.31 & 2400 &  2700 & 0.92 & & 0.00 & $-$ & $-$ & 2.5  & 176.2 & 28.41 &   \phantom{0}1 $-$   10 \\
2003~BM$_{4}$     & Earth      & 1995 Jul 29.158 & 0.01 &15400 & 20400 & 0.99 & & 0.00 & $-$ & $-$ & 33.0 & 179.8 & 24.68 &             20 $-$   60 \\
1998~BT$_{13}$    & Earth      & 1998 Jan 23.374 & 1.00 & 3500 &  3600 & 0.93 & & 0.00 & $-$ & $-$ & 6.9  & 179.6 & 26.36 &             20          \\
1998~BT$_{13}$    & Moon       & 1998 Jan 23.915 & 1.00 & 3900 &  4000 & 0.98 & & 0.00 & $-$ & $-$ & 7.2  & 173.2 & 26.36 &             20          \\
1998~DX$_{11}$    & Moon       & 1998 Feb 23.615 & 1.00 & 2800 &  3000 & 0.98 & & 0.00 & $-$ & $-$ & 5.5  & 173.0 & 27.06 &             10 $-$   20 \\
1999~HC$_{1}$     & Earth      & 1999 Apr 18.970 & 1.00 & 6900 &  7400 & 0.79 & & 0.00 & $-$ & $-$ & 8.7  & 179.7 & 24.48 &             30 $-$   80 \\
1999~TM$_{13}$    & Earth      & 1999 Oct  3.950 & 1.00 &31000 & 31100 & 0.98 & & 0.00 & $-$ & $-$ & 31.0 & 179.8 & 23.58 &             30 $-$   70 \\
2003~YT$_{70}$    & Earth      & 1999 Dec  6.974 & 0.01 &42000 & 50700 & 0.97 & & 0.00 & $-$ & $-$ & 21.8 & 179.8 & 25.55 &             20 $-$   40 \\
2001~UC$_{5}$     & Earth      & 2001 Oct 21.724 & 1.00 & 4400 &  4600 & 0.98 & & 0.00 & $-$ & $-$ & 24.2 & 179.8 & 21.32 &            160 $-$  370 \\
2003~YT$_{70}$    & Earth      & 2001 Dec  6.699 & 0.03 &39600 & 50100 & 0.97 & & 0.00 & $-$ & $-$ & 22.2 & 179.8 & 25.55 &             20 $-$   40 \\
2002~DQ$_{3}$     & Earth      & 2002 Mar  1.482 & 1.00 &29400 & 29500 & 0.97 & & 0.00 & $-$ & $-$ & 21.6 & 179.8 & 23.77 &  \phantom{1}50 $-$  110 \\
2002~VY$_{91}$    & Moon       & 2002 Nov 11.440 & 1.00 & 4300 &  4400 & 0.98 & & 0.00 & $-$ & $-$ & 7.4  & 171.8 & 26.03 &             10 $-$   30 \\
2005~ES$_{70}$    & Earth      & 2003 Mar 14.076 & 0.02 & 5600 &  6600 & 0.97 & & 0.00 & $-$ & $-$ & 21.3 & 179.8 & 23.56 &  \phantom{1}50 $-$  120 \\
2003~LW$_{1}$     & Earth      & 2003 Jun  6.394 & 1.00 & 4900 &  5000 & 0.97 & & 0.00 & $-$ & $-$ & 16.8 & 179.7 & 23.13 &  \phantom{1}60 $-$  150 \\
2003~SY$_{4}$     & Earth      & 2003 Sep 17.296 & 1.00 & 2100 &  2100 & 0.98 & & 0.00 & $-$ & $-$ & 6.2  & 179.6 & 26.54 &             10 $-$   30 \\
2003~UR$_{25}$    & Earth      & 2003 Oct 17.634 & 0.08 &28700 & 38800 & 0.98 & & 0.00 & $-$ & $-$ & 32.3 & 179.8 & 25.61 &             20 $-$   50 \\
2003~UT$_{55}$    & Moon       & 2003 Oct 26.746 & 1.00 & 2000 &  2100 & 0.94 & & 0.00 & $-$ & $-$ & 3.1  & 174.8 & 27.36 &             10 $-$   20 \\
2003~WT$_{153}$   & Earth      & 2003 Nov 28.364 & 1.00 & 7800 &  7800 & 0.25 & & 0.00 & $-$ & $-$ & 3.0  & 179.7 & 28.05 &   \phantom{0}1 $-$   10 \\
2003~YH$_{111}$   & Earth      & 2003 Dec 24.350 & 1.00 &14700 & 15600 & 0.98 & & 0.00 & $-$ & $-$ & 18.2 & 179.7 & 24.54 &             30 $-$   80 \\
2004~HD           & Earth      & 2004 Apr 25.136 & 1.00 &12200 & 12200 & 0.94 & & 0.00 & $-$ & $-$ & 15.3 & 179.8 & 24.65 &             30 $-$   70 \\
2004~SR$_{26}$    & Moon       & 2004 Sep 20.964 & 1.00 & 2900 &  3200 & 0.95 & & 0.00 & $-$ & $-$ & 4.3  & 166.8 & 26.43 &             10 $-$   30 \\
2004~ST$_{26}$    & Moon       & 2004 Sep 22.089 & 0.18 &  800 &  1200 & 0.00 & & 0.04 & 400 & 500 & 1.1  & 117.2 & 26.37 &             10 $-$   30 \\
2001~SE$_{270}$   & Earth      & 2005 Sep 20.072 & 0.02 &16500 & 20800 & 0.99 & & 0.00 & $-$ & $-$ & 33.8 & 179.8 & 25.10 &             20 $-$   60 \\  
2003~YT$_{70}$    & Earth      & 2005 Dec  6.364 & 0.03 &41000 & 51700 & 0.97 & & 0.00 & $-$ & $-$ & 22.2 & 179.8 & 25.55 &             20 $-$   40 \\
2004~YG$_{1}$     & Earth      & 2005 Dec 23.741 & 0.01 & 5800 &  8900 & 0.98 & & 0.00 & $-$ & $-$ & 26.8 & 179.7 & 21.35 &            140 $-$  320 \\
2000~TH$_{1}$     & Earth      & 2007 Oct  6.537 & 0.01 &13200 & 17600 & 0.99 & & 0.00 & $-$ & $-$ & 36.1 & 179.8 & 22.36 &  \phantom{1}90 $-$  220 \\
2003~YT$_{70}$    & Earth      & 2007 Dec  6.345 & 0.01 &34500 & 51700 & 0.97 & & 0.00 & $-$ & $-$ & 22.2 & 179.8 & 25.55 &             20 $-$   40 \\
1998~SD$_{9}$     & Earth      & 2008 Sep  9.407 & 0.02 & 5900 &  7000 & 0.95 & & 0.00 & $-$ & $-$ & 17.0 & 179.8 & 24.15 &  \phantom{1}40 $-$  100 \\
1998~VF$_{32}$    & Earth      & 2009 Nov 18.679 & 0.02 & 4700 &  6000 & 0.97 & & 0.00 & $-$ & $-$ & 21.0 & 179.8 & 21.15 &            180 $-$  400 \\
2001~FE$_{90}$    & Earth      & 2009 Jun 30.402 & 0.01 & 4700 &  5900 & 0.80 & & 0.00 & $-$ & $-$ & 8.4  & 179.8 & 19.81 &            330 $-$  730 \\
2001~FE$_{90}$    & Moon       & 2009 Jun 30.632 & 0.02 & 4200 &  4800 & 0.99 & & 0.00 & $-$ & $-$ & 8.6  & 173.3 & 19.81 &            330 $-$  730 \\
2003~YT$_{70}$    & Earth      & 2009 Dec  6.178 & 0.01 &41400 & 53000 & 0.97 & & 0.00 & $-$ & $-$ & 22.9 & 179.8 & 25.55 &             20 $-$   40 \\
2000~TU$_{28}$    & Earth      & 2010 Oct 16.730 & 0.02 &12300 & 13200 & 0.98 & & 0.00 & $-$ & $-$ & 30.0 & 179.8 & 20.42 &            190 $-$  420 \\
2001~LB           & Earth      & 2012 Jun 11.647 & 0.01 & 9600 & 10800 & 0.98 & & 0.00 & $-$ & $-$ & 30.6 & 179.8 & 20.77 &            200 $-$  460 \\
2004~YA$_{5}$     & Earth      & 2012 Dec 21.583 & 0.01 & 4600 &  7300 & 0.99 & & 0.00 & $-$ & $-$ & 31.2 & 179.8 & 22.58 &  \phantom{1}90 $-$  200 \\
1997~WQ$_{23}$    & Earth      & 2013 Nov 18.320 & 0.01 &61000 & 70000 & 0.99 & & 0.00 & $-$ & $-$ & 34.1 & 179.8 & 20.39 &            230 $-$  530 \\
2004~BL$_{86}$    & Earth      & 2015 Jan 27.205 & 0.04 & 1500 &  1900 & 0.05 & & 0.00 & $-$ & $-$ & 3.6  & 179.6 & 18.88 & \phantom{1}500 $-$ 1100 \\
2004~BL$_{86}$    & Moon       & 2015 Jan 27.337 & 0.03 &  900 &  1200 & 0.93 & & 0.00 & $-$ & $-$ & 3.7  & 164.4 & 18.88 & \phantom{1}500 $-$ 1100 \\
2002~LY$_{1}$     & Earth      & 2016 May 29.100 & 0.13 &14300 & 18300 & 0.99 & & 0.00 & $-$ & $-$ & 38.2 & 179.8 & 21.92 &            120 $-$  280 \\
2004~SS           & Earth      & 2017 Sep 24.522 & 0.01 & 6300 &  9100 & 0.92 & & 0.00 & $-$ & $-$ & 13.5 & 179.7 & 21.95 &            110 $-$  260 \\
2002~CB$_{19}$    & Earth      & 2018 Feb  7.188 & 0.02 & 6000 &  8300 & 0.96 & & 0.00 & $-$ & $-$ & 17.8 & 179.8 & 24.76 &             30 $-$  60  \\
1999~LK$_{1}$     & Earth      & 2018 May 21.361 & 0.09 & 8400 & 10100 & 0.96 & & 0.00 & $-$ & $-$ & 19.8 & 179.8 & 22.11 &            110 $-$ 250  \\
2001~GP$_{2}$     & Earth      & 2020 Oct  6.662 & 0.01 & 8600 &  9500 & 0.00 & & 0.01 &4200 &5000 &  1.1 & 179.3 & 26.88 &             10 $-$  20  \\
2001~GP$_{2}$     & Moon       & 2020 Oct  7.075 & 0.01 & 3800 &  3900 & 0.78 & & 0.00 & $-$ & $-$ &  1.8 & 151.0 & 26.88 &             10 $-$  20  \\
2003~AF$_{23}$    & Earth      & 2021 Jan  7.156 & 0.03 & 4800 &  6100 & 0.96 & & 0.00 & $-$ & $-$ & 19.2 & 179.8 & 20.58 &            210 $-$ 480  \\
2004~FH           & Earth      & 2021 Feb 20.918 & 0.96 & 3800 &  9200 & 0.98 & & 0.00 & $-$ & $-$ & 25.0 & 179.7 & 26.39 &             20 $-$  40  \\
2000~PN$_{8}$     & Earth      & 2021 Aug 15.655 & 0.01 & 8700 & 11000 & 0.99 & & 0.00 & $-$ & $-$ & 35.0 & 179.8 & 22.13 &            100 $-$ 240  \\
1999~FJ$_{21}$    & Earth      & 2021 Oct 12.873 & 0.02 & 7200 &  9200 & 0.98 & & 0.00 & $-$ & $-$ & 27.0 & 179.8 & 20.59 &            220 $-$ 510  \\
23606             & Earth      & 2022 Jul 19.642 & 1.00 &15900 & 16600 & 0.99 & & 0.00 & $-$ & $-$ & 39.8 & 179.8 & 18.47 & \phantom{1}600 $-$ 1300 \\
2004~UT$_{1}$     & Moon       & 2022 Oct 25.182 & 0.05 & 9800 & 12300 & 0.98 & & 0.00 & $-$ & $-$ & 6.7  & 179.4 & 26.43 &             10 $-$   30 \\
2004~UT$_{1}$     & Earth      & 2022 Oct 25.238 & 0.06 & 9400 & 11300 & 0.70 & & 0.00 & $-$ & $-$ & 6.6  & 179.7 & 26.43 &             10 $-$   30 \\
2001~LB           & Earth      & 2023 Jun 12.402 & 0.01 &10300 & 10800 & 0.98 & & 0.00 & $-$ & $-$ & 30.6 & 179.8 & 20.77 &            200 $-$  460 \\
2000~YS$_{134}$   & Earth      & 2023 Dec 29.199 & 0.15 &11100 & 14800 & 0.97 & & 0.00 & $-$ & $-$ & 23.3 & 179.8 & 23.24 &  \phantom{1}70 $-$  160 \\
2005~FG           & Earth      & 2024 Mar 16.098 & 0.01 &17100 & 28100 & 0.99 & & 0.00 & $-$ & $-$ & 33.3 & 179.8 & 24.00 &  \phantom{1}50 $-$  110 \\
99942             & Moon       & 2029 Apr 14.132 & 0.74 & 1900 &  2500 & 0.00 & & 0.06 & 250 & 350 & 0.35 &  17.1 & 19.17 &            430 $-$  970 \\
2003~AF$_{23}$    & Earth      & 2030 Jan  7.410 & 0.02 & 4400 &  6100 & 0.96 & & 0.00 & $-$ & $-$ & 19.1 & 179.8 & 20.58 &            210 $-$  480 \\
2005~CD$_{69}$    & Earth      & 2030 Feb 25.433 & 0.01 &30300 & 30800 & 0.95 & & 0.00 & $-$ & $-$ & 17.0 & 179.8 & 24.11 &             40 $-$   90 \\
2003~WT$_{153}$   & Earth      & 2030 Nov 17.881 & 0.34 &12700 & 15000 & 0.91 & & 0.00 & $-$ & $-$ & 12.9 & 179.8 & 28.05 &   \phantom{0}1 $-$   10 \\
2005~BC           & Earth      & 2034 Jul 15.295 & 0.01 & 4000 &  5200 & 0.96 & & 0.00 & $-$ & $-$ & 20.0 & 179.8 & 18.12 & \phantom{1}700 $-$ 1500 \\
2002~AB$_{2}$     & Earth      & 2039 Jan  3.978 & 0.01 &10300 & 12200 & 0.99 & & 0.00 & $-$ & $-$ & 35.1 & 179.8 & 22.99 &  \phantom{1}60 $-$  140 \\
2004~ER$_{21}$    & Earth      & 2039 Mar 18.246 & 0.04 &10900 & 13500 & 0.97 & & 0.00 & $-$ & $-$ & 20.1 & 179.8 & 24.29 &             30 $-$   80 \\
1999~VX$_{25}$    & Moon       & 2040 Nov 13.616 & 0.01 & 7800 &  8600 & 0.98 & & 0.00 & $-$ & $-$ & 8.0  & 173.6 & 26.70 &             10 $-$   30 \\
1999~VX$_{25}$    & Earth      & 2040 Nov 14.063 & 0.01 &11400 & 12500 & 0.78 & & 0.00 & $-$ & $-$ & 7.9  & 179.7 & 26.70 &             10 $-$   30 \\
2005~KA           & Moon       & 2042 May  4.445 & 0.01 & 2900 &  3300 & 0.93 & & 0.00 & $-$ & $-$ & 4.7  & 178.3 & 24.73 &             30 $-$   70 \\
2005~KA           & Earth      & 2042 May  5.035 & 0.01 & 5600 &  7300 & 0.40 & & 0.00 & $-$ & $-$ & 4.7  & 179.7 & 24.73 &             30 $-$   70 \\
1997~YM$_{9}$     & Earth      & 2044 Dec 26.292 & 0.01 &12300 & 14100 & 0.93 & & 0.00 & $-$ & $-$ & 13.5 & 179.8 & 24.77 &             30 $-$   60 \\
2000~LF$_{3}$     & Earth      & 2046 Jun 12.814 & 0.01 & 2600 &  2800 & 0.27 & & 0.00 & $-$ & $-$ & 4.3  & 179.7 & 21.57 &            140 $-$  320 \\ 
2004~DK$_{1}$     & Earth      & 2045 Apr 29.258 & 0.01 &65200 & 81800 & 0.99 & & 0.00 & $-$ & $-$ & 33.3 & 179.8 & 21.05 &            170 $-$  380 \\
2001~QJ$_{142}$   & Earth      & 2047 Sep 26.214 & 0.03 &25500 & 29500 & 0.99 & & 0.00 & $-$ & $-$ & 37.3 & 179.8 & 23.42 &  \phantom{1}60 $-$  130 \\
\enddata
\tablecomments{Near Earth Asteroids eclipsed by Earth and Moon
in the period 1990-2050, sorted by epoch.
Probability $P$, mean and 
maximum time $T$, averaged over the VAs experiencing the eclipse,
are reported for penumbra and umbra eclipses.
The average Earth distance $d_E$ and Solar elongation $\lambda$ are also showed.
For penumbra eclipses, the minimum Solar Disk Fraction $\Gamma$ is also provided.
The eclipse epoch is referred to the minimum of $\Gamma$ in case of a partial eclipse,
or to the central moment for a total eclipse.
Asteroid diameters are obtained from the Near-Earth Asteroids Data Base,
European Asteroid Research Node (\url{http://earn.dlr.de/}).
The absolute magnitude $H$ is obtained from the NEODyS server \citep{neodys}.
This list is limited to eclipses within 40~LD from the Earth and 10~LD from the Moon,
with $P_p \ge$ 0.01 and
min($\Gamma$) $\le$ 0.99.
}
\end{deluxetable}

%% file: immediate_table.tex
\begin{deluxetable}{llllr} 
\tablecolumns{5}
\tablewidth{0pc}
\tablecaption{Immediate Eclipses.\label{immediate_table}}
\tablehead{
\colhead{asteroid} & 
\colhead{object} &
\colhead{discovery epoch} &
\colhead{eclipse epoch} &
\colhead{$\Delta$t}
\\
\colhead{} & 
\colhead{} & 
\colhead{[TDT]} &
\colhead{[TDT]} &
\colhead{[day]}
}
\startdata
1991~TU           & Moon       & 1991 Oct  7.310 & 1991 Oct  7.279 & -0.031 \\
1998~BT$_{13}$    & Earth      & 1998 Jan 24.403 & 1998 Jan 23.374 & -1.029 \\
1998~BT$_{13}$    & Moon       & 1998 Jan 24.403 & 1998 Jan 23.915 & -0.488 \\
1998~DX$_{11}$    & Moon       & 1998 Feb 23.352 & 1998 Feb 23.615 &  0.263 \\ 
1999~HC$_{1}$     & Earth      & 1999 Apr 16.311 & 1999 Apr 18.970 &  2.659 \\ 
1999~TM$_{13}$    & Earth      & 1999 Oct  5.287 & 1999 Oct  3.950 & -1.337 \\
2001~UC$_{5}$     & Earth      & 2001 Oct 20.251 & 2001 Oct 21.724 &  1.473 \\ 
2002~DQ$_{3}$     & Earth      & 2002 Feb 22.405 & 2002 Mar  1.482 &  7.077 \\ 
2002~VY$_{91}$    & Moon       & 2002 Nov 12.227 & 2002 Nov 11.440 & -0.787 \\ 
2003~LW$_{1}$     & Earth      & 2003 Jun  3.308 & 2003 Jun  6.394 &  3.086 \\ 
2003~SY$_{4}$     & Earth      & 2003 Sep 17.291 & 2003 Sep 17.296 &  0.005 \\ 
2003~UR$_{25}$    & Earth      & 2003 Oct 25.200 & 2003 Oct 17.634 & -7.566 \\
2003~UT$_{55}$    & Moon       & 2003 Oct 26.938 & 2003 Oct 26.746 & -0.192 \\
2003~WT$_{153}$   & Earth      & 2003 Nov 29.315 & 2003 Nov 28.364 & -0.951 \\
2003~YH$_{111}$   & Earth      & 2003 Dec 27.301 & 2003 Dec 24.350 & -2.951 \\
2004~HD           & Earth      & 2004 Apr 16.199 & 2004 Apr 25.136 &  8.937 \\ 
2004~SR$_{26}$    & Moon       & 2004 Sep 22.234 & 2004 Sep 20.964 & -1.270 \\
2004~ST$_{26}$    & Moon       & 2004 Sep 23.285 & 2004 Sep 22.089 & -1.196 \\
\enddata
\tablecomments{Near Earth Asteroids experiencing an eclipse during the
same close approach relative to their discovery. See Section~\ref{sec_imm_ecl}
and Figure~\ref{fig_mag_delayeclipse} for more details.
The case of 2003~SY$_{4}$ is slightly different from all the
other eclipses in this table:
part of the discovery observations were made during the eclipse, see Section~\ref{sec_SY}.
}
\end{deluxetable}